\documentclass[a4paper]{jpconf}
\usepackage{graphicx}
\usepackage{multicol}
\usepackage{wrapfig}
\usepackage{capt-of}

\begin{document}
\title{Calibrating Analytical Models for Semilocal Strings}

\author{A. Lopez-Eiguren\footnote{Work in collaboration with A.Ach\'ucarro, A.Avgoustidis, A.M.M.Leite, C.J.A.P.Martins, A.S.Nunes, J.Urrestilla}}%$^1$,A. Ach\'ucarro$^{1,2}$, A. Avgoustidis$^3$, A. M. M. Leite$^{4,5}$, C. J. A. P. Martins$^6$, A. S. Nunes$^{7,8}$, J. Urrestilla$^1$}
\address{Department of Theoretical Physics, University of the Basque Country UPV-EHU, 48040 Bilbao, Spain}
%\address{$^2$ Institute Lorentz of Theoretical Physics, University of Leiden, 2333CA Leiden, The Netherlands}
%\address{$^3$ School of Physics and Astronomy, University of Nottingham, University Park, Nottingham NG7 2RD, England}
%\address{$^4$ Centro de Astrof\'{\i}sica, Universidade do Porto, Rua das Estrelas, 4150-762 Porto, Portugal}
%\address{$^5$ \'Ecole Polytechnique, 91128 Palaiseau Cedex, France}
%\address{$^6$ Centro de Astrof\'{\i}sica, Universidade do Porto, Rua das Estrelas, 4150-762 Porto, Portugal}
%\address{$^7$ Centro de Astrof\'{\i}sica, Universidade do Porto, Rua das Estrelas, 4150-762 Porto, Portugal}
%\address{$^8$ Faculdade de Ci\^encias da Universidade de Lisboa, Campo Grande, 1749-016 Lisboa, Portugal}

\ead{asier.lopez@ehu.es}%, ana, tavgoust@gmail.com, up080322016@alunos.fc.up.pt, Carlos.Martins@astro.up.pt, up200807913@fc.up.pt, jon.urrestilla@ehu.es}

\begin{abstract}
In this work we calibrate two different analytic models of semilocal strings by constraining the values of their free parameters. In order to do so, we  use data obtained from  the largest and most accurate field theory simulations of semilocal strings to date, and compare several key properties with the predictions of the models. As this is still work in progress, we present some preliminary results together with descriptions of the methodology we are using in the  characterisation of semilocal string networks.
  
\end{abstract}

\section{Introduction}

Understanding the evolution of string networks is crucial for predicting their number densities, which in turn determine their potentially observable effects. However, the quantitatively accurate modelling of string network evolution is a difficult problem, requiring the combination of a range of techniques (both numerical and analytical), and interpolating between physics at very different energy scales.

Here, we present a calibration of analytical models for semilocal strings using field theory simulations. 
This work is part of an ongoing project where we tackle in turn different aspects of the calibration procedure by comparing the numerical simulations with predictions for the analytic models. Firstly, in section  \ref{lsp},  we study the large-scale properties of the simulated networks  \cite{Achucarro:2013mga}.
Then, in section~\ref{sd}, we outline  the comparison between field theory simulations and the analytic models, anticipating  work that will appear in \cite{Achucarro:2014mga}. Section~\ref{v}  shows the last ingredient of our analysis, where we present the techniques we will use in a future work 
 \cite{Achucarro:2015mga} to estimate the velocities of the semilocal strings. Prior to all this, in section~\ref{m} we introduce the semilocal model and the analytic models used in  this work, as well as the numerical simulations performed.

\section{The Model \label{m}}
\subsection{Semilocal Model \label{slm}}
Semilocal strings \cite{Vachaspati:1991dz,Achucarro:1999it,Hindmarsh:1991jq} were introduced as a minimal extension of the Abelian Higgs (AH) model with two complex scalar fields instead of just one, that form an $SU(2)$ doublet. This leads to $U(1)$ flux-tube solutions even though the vacuum manifold is simply connected. The strings of this extended model have some similarities with ordinary local $U(1)$ strings, but they are not purely topological and will therefore have different properties. For example, since they are not topological, they need not be closed or infinite and can have ends. These ends are effectively global monopoles which have long-range interactions\cite{Hindmarsh:1992yy}. %The monopoles at the ends of the strings have some exotic properties by themselves \cite{Achucarro:2000td}.

The relevant action for the simplest semilocal string model, the one we will use in the numerical simulations, reads:
\begin{equation}
S= \int d^4 x \Big( \big[ (\partial_{\mu}-i A_{\mu})\Phi \big]^2 - \frac{1}{4}F^2-\frac{\beta}{4}(\Phi^+ \Phi)^2 \Big),
\label{action}
\end{equation}
where $\Phi=(\phi,\varphi)$ ($\phi$ and $\varphi$ are complex scalar fields), $F^2=F_{\mu\nu}F^{\mu\nu}$ and $F_{\mu\nu}=\partial_{\mu}A_{\nu}-\partial_{\nu}A_{\mu}$. It can be easily seen that setting one of the two scalar fields to zero, we recover the AH model.

The stability of the strings is not trivial, and it will depend on the value of the parameter $\beta=m_{scalar}^2/m_{gauge}^2$: for $\beta<1$ the string is stable, for $\beta>1$ it is unstable, and for $\beta=1$ it is neutrally stable \cite{Vachaspati:1991dz,Achucarro:1999it,Hindmarsh:1991jq}. Only low values of $\beta$ will be of interest for the comparison, because otherwise the string network is either unstable or disappears very fast. It is also known that the lower the $\beta$, the more it looks like AH strings.
\subsection{Analytic Models}

Broadly speaking there are two ways to model the evolution of these defects \cite{Nunes:2011sf} using effective models. The simplest one attempts to model the overall network, specifically focusing on the evolution of the monopoles (which are known to dominate the dynamics of the network). This is a 'classical' one-scale model description. On the other hand, given that the evolution of the semilocal segments is highly non-trivial, a more accurate description must necessary include the evolution of this segment population. In this subsection we will concentrate on the latter.

Numerical simulations indicate that while some segments shrink and disappear, there is also a significant probability that segments merge and  form longer segments. This is due to the long range interactions of the global monopoles at the end of  segments. To account for this behaviour we start from the evolution equations for segment size introduced in the velocity-dependent one-scale (VOS) model, where the equations come from the statistical averaging of the microscopic equations of motion \cite{Martins:1996jp,Martins:2000cs}:

\begin{equation}
\frac{dl_s}{dt}=Hl_s-v_s^2\frac{l_s}{l_d}, \;\;\;\; \frac{dv_s}{dt}=(1-v_s^2)\Big[ \frac{k}{l_s}-\frac{v_s}{l_d}\Big],
\label{vos}
\end{equation}
where $l_s$ is the length of the segment under consideration, $v_s$ its the root mean square (RMS) velocity, $k$ a free parameter describing string curvature(to be calibrated), H the Hubble parameter and $l_d$ is the string damping length. We now modify those equations in two possible phenomenological ways \cite{Nunes:2011sf}:

\subsubsection{Scale-dependent Behaviour: \label{A}}
A simple generalisation of the equations (\ref{vos}) would be
\begin{equation}
\frac{dl_s}{dt}=Hl_s-v_s^2\frac{l_s}{l_d}+\sigma \Big(1-\frac{L}{l_s}\Big) v_m^2
\label{A-1}
\end{equation}
\begin{equation}
\frac{dv_s}{dt}=(1-v_s^2)\Big[\frac{k}{l_s}-\frac{v_s}{l_d}\Big]
\label{A-2}
\end{equation}
where $L$ is the characteristic scale of the monopoles and $\sigma$ is a free parameter controlling the importance of the newly introduced term. The new term was added on phenomenological reasoning that, to a first approximation, small segments should shrink and large ones should grow and merge \cite{Hindmarsh:1992yy}. This can be intuited as a competition between two characteristic timescales. Each segment will have an annihilation timescale, and each monopole will have a characteristic timescale in which to find its (anti)partner and annihilate, thereby producing a longer segment. The second process is expected to become relatively more likely as the segment size increases.

\subsubsection{Balance Equation: \label{B}}
The following modification for the evolution equations is considered

\begin{equation}
\frac{dl_s}{dt}=Hl_s-v_s^2\frac{l_s}{l_d}+dv_s \Big(  \frac{l_s}{L}-1 \Big)
\label{B-1}
\end{equation}
\begin{equation}
\frac{dv_s}{dt}=(1-v_s^2)\Big[\frac{k}{l_s}-\frac{v_s}{l_d}\Big]
\label{B-2}
\end{equation}
In these we are assuming that the network of string segments has a Brownian distribution, something that can be tested in numerical simulations. The new term (including new free parameter, d) accounts for the probability that different segments intersect, which depends both on the length/number density and velocity of the segments.

\subsection{Field Theory Simulations}
We simulated numerically the semilocal model introduced in section \ref{slm} so as to provide us with data to be used for comparison with the analytic models. The parameter space we want to explore is rather large, so we carefully chose the cases to study, and tried to maximise the information we could obtain from simulations given the computer resources available to us. The cases for study chosen are $\beta=0.01,0.04,0.09$; the cosmologies under consideration correspond to expanding universes, both in radiation and matter eras. We perform each one of these cases for two different lattice spacings $\delta x=0.5,1$.

We discretised the action given in equation (\ref{action}) by standard techniques (using lattice-link variables and a staggered-leapfrog method) and evolved the discretised action in $1024^3$ lattices with periodic boundary conditions.

Once the system reaches scaling, quantities of interest can be measured. Semilocal strings are non topological entities; therefore, we cannot use topology to detect them. This kind of strings can be thought of as concentrations of magnetic energy, and that is the strategy we follow: we first calculate the maximum of the magnetic field strength, and the radius, of a straight and infinite AH string for a given $\beta$. We use those values for the simulated semilocal string network: if the magnetic field strength of a simulated semilocal model measured at a point of the box exceeds the $25\%$ of the maximum of the corresponding AH string, we consider that point to be part of a semilocal string segment. The output of our simulation is thus an array of points from the simulated box which have a considerable concentration of magnetic field strength.
\begin{figure}[h]
\begin{minipage}{18pc}
\includegraphics[width=15.5pc]{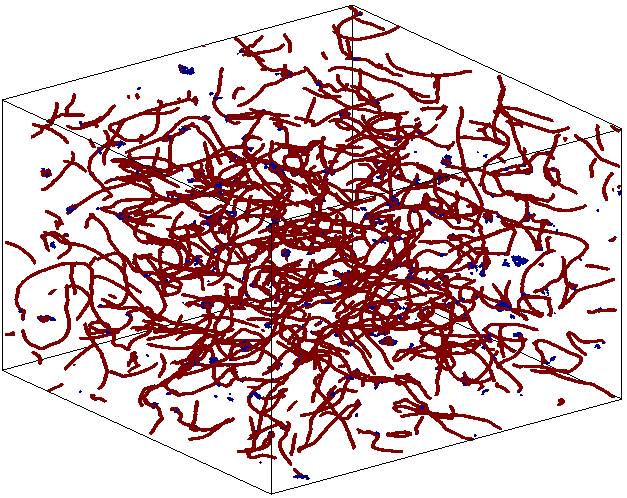}
\end{minipage}\hspace{2pc}%
\begin{minipage}{16pc}
\includegraphics[width=15.5pc]{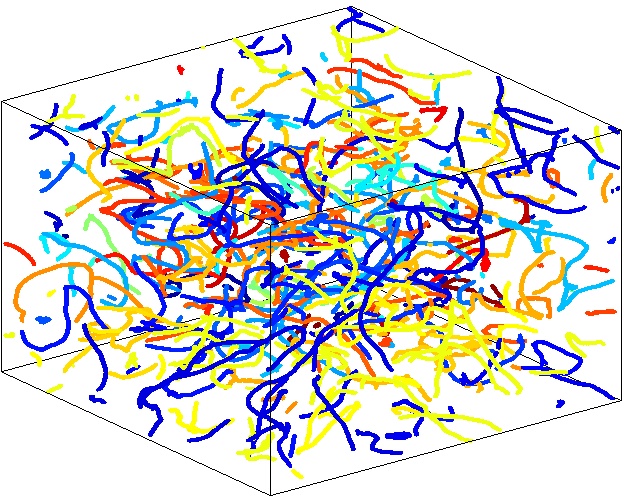}
\end{minipage} 
\caption{\label{box} Semilocal string network, in matter domination with $\beta = 0.04$. The left figure shows two types of structures: on the one hand we have tube-like structures (proper strings) and on the other short blobs. These blobs we disregard in our analysis. The right figure shows the network without blobs, and also each segment has been identified and plotted with a different colour. As the number of segments is large, the colours are unfortunately used for more than one string segment. Note also that the blob removal procedure does sometimes fail to identify some sphere-like structures, since their volume is large.}
\end{figure}

We then group together the points that have been output by the simulations into segments. These segments are mostly tube-like, but some are sphere-like instead of tube-like, i.e., they are blobs of energy. These can be formed, for example, after a segment collapsed into itself. We do not wish to count these blobs as part of our network, and we introduce a lower cut-off: those segments that are not longer than a given factor times the typical radius of a string are considered to be blobs and are discarded. One typical simulation snapshot is shown in Fig.~\ref{box}.

\section{Large-scale Properties \cite{Achucarro:2013mga} \label{lsp}}

We have performed 12 simulations for each case of our set of parameters, and use the results to obtain basic statistics about the properties of the networks. All in all, for each one of those simulations, and for specific values of the simulation time, we obtain the total string length ${\cal L}(t)$ and monopole number ${\cal N}(t)$ in the box. Both of these provide simple diagnostics for the large-scale evolution of the network, and specifically for the presence of scaling.

Fig. \ref{scaling} provides an example of the evolution of these quantities for the case $\beta=0.09$, $\delta x=1$ and matter era simulation. This is representative of all the sets of simulations we have performed. This analysis therefore shows that all the networks have reached the scaling solution.

The string lengths and number of monopoles obtained can easily be translated into VOS-type length scales using \cite{Martins:1996jp,Martins:2000cs},
\begin{multicols}{2}

 \begin{equation}
 \gamma_s \equiv \frac{L_s}{t}=\frac{1}{t}\sqrt{\frac{V}{\cal L}},
 \end{equation}

\begin{equation}
\gamma_m \equiv \frac{L_m}{t}=\frac{1}{t}\Big(\frac{V}{\cal N}\Big)^{1/3}.
\end{equation}

After analysing the VOS-type length scales, in all cases under study, we can see that in matter era $\gamma_s$ is larger and $\gamma_m$ smaller than in radiation era. This is because in radiation era the effective monopole velocity is higher and segments move faster to either grow and meet with other segments or collapse, giving a longer typical string length and smaller number of monopoles. Note that $\gamma_s$ and $\gamma_m$ are inversely proportional to ${\cal L}$ and ${\cal N}$, respectively.

\begin{minipage}{\linewidth}

\includegraphics[width=18pc]{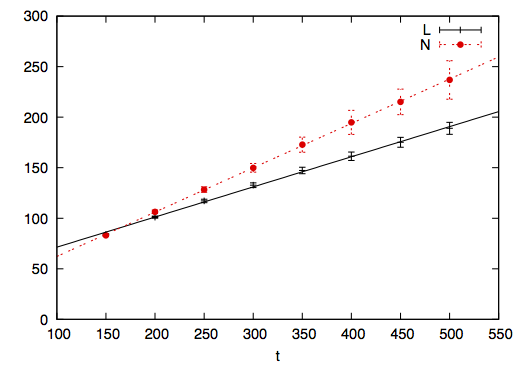}
\captionof{figure}{Scaling plots for L and N for $\beta=0.09$, in the matter era, with $\delta x = 1$. The error bars show statistical errors over the 12 simulations.}
\label{scaling}
\end{minipage}
\end{multicols}

However for a given cosmology, $\gamma_s$ grows with $\beta$ and $\gamma_m$ gets smaller. This is also to be expected since for lower $\beta$ we expect the system to behave more like an AH network, which has longer strings and fewer segments.

\section{Segment Distribution \cite{Achucarro:2014mga} \label{sd}}

Our simulations have sufficient resolution and dynamical range to allow us to perform quantitative diagnostics, and study in much more detail the structure and evolution of our string networks. While the previous section focused only on two important physical quantities, ${\cal L}$ and ${\cal N}$, in this section we will take this further and study the statistics of string segment length distributions. Starting from initial configurations of the semilocal string networks, we can group, at any given time, all string segments into suitable chosen length bins and study how the length distributions evolve in time.

\begin{figure}[h]
\begin{minipage}{20pc}
\includegraphics[width=22pc]{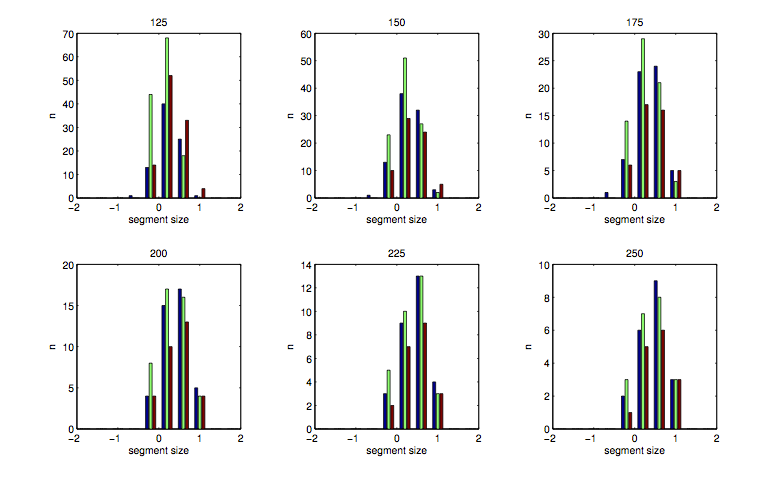}
\caption{\label{sigma-evo} Number of strings in each bin for different time steps for $\beta=0.01$, radiation era and $\delta x=0.5$. The red columns are the values coming directly from the simulations, while blue columns represents the \ref{A} model and green columns the \ref{B} model.}
\end{minipage}\hspace{2pc}%
\begin{minipage}{18pc}
\includegraphics[width=18pc]{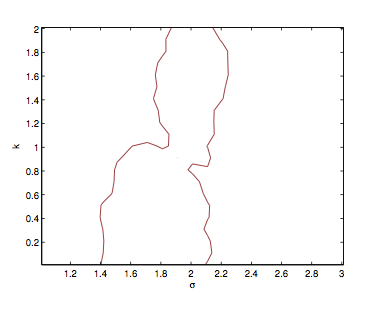}
\caption{\label{sigma-contour}Contour plot showing the best values for the parameter space for $\beta=0.01$, radiation era and $\delta x=0.5$.}
\end{minipage} 
\end{figure}

Furthermore we can evolve these length distributions using our analytic models, so as to compare data from field theory simulations and analytic models. For example in Fig. \ref{sigma-evo} we can see length distributions for different time steps directly coming from the simulations, as well as, from analytic models. Now we can employ $\chi^2$ minimisation to calibrate the free parameters of the analytic models.

In Fig \ref{sigma-contour} we have a typical contour plot obtained after the $\chi^2$ minimisation. The figure shown is a preliminary result for the scale-dependent behaviour case. There we can see that the parameter $\sigma$ is constrained $[1.4,2.2]$, but the parameter $k$ is unconstrained. 

\section{Velocities \cite{Achucarro:2015mga} \label{v}}

In this last section we will reduce systematic uncertainties in our modelling to give a fully quantitative calibration. To do this, we will calculate transverse velocity of the string segments and also we will measure the monopole velocity by tracking monopoles during the evolution. The results obtained using these method will appear in \cite{Achucarro:2015mga}.

\subsection{String Transverse Velocity}

Firstly in order to work out the transverse velocities of the strings we will use the method proposed by \cite{Hindmarsh:2008dw}. 

\begin{equation}
\gamma_F^2 \langle \dot{\bf x}^2 \rangle _{F}=\frac{{\bf E}^2_{\cal L}}{{\bf B}^2_{\cal L}} \;\;\;\;\;\;\; \gamma_G^2 \langle \dot{\bf x}^2 \rangle_{G} =\frac{{\bf \Pi}^2_{\cal L}}{({\bf D}\Phi)^2_{\cal L}}
\end{equation}
In the first case $E_i=F_{0i}$ and $B_k=\epsilon_{ijk}F_{ij}/2$ are the electric and magnetic fields respectively where $\epsilon_{ijk}$ is the Levi-Civita symbol. In the second case $\Pi=\dot{\Phi}$ is the canonical momentum and ${\bf D}\Phi=\partial \Phi/\partial {\bf x} +i {\bf A} \Phi$ represents the spatial gradients of the fields. $\gamma_F^2$ and $\gamma_G^2$ are calculated using $\langle \dot{\bf x}^2 \rangle_F$ and $\langle \dot{\bf x}^2 \rangle_G$ respectively, and subscript ${\cal L}$ denotes a Lagrangian weighting of a field ${\cal X}$ according to

\begin{equation}
{\cal X}_{\cal L}=\frac{\int d^3 x \; {\cal X} \;{\cal L}}{\int d^3x \; {\cal L}}.
\end{equation}

\subsection{Monopole Velocity}

As said before semilocal strings can have ends. These ends are effectively global monopoles, and we investigate the velocity of these ends. In order to measure this velocity, we need to pinpoint where a monopole moves at every step, but it is difficult taking into account that there are lots of strings in the simulation box. 

Firstly, we have to detect all the monopoles individually. As we know, monopoles are characterised by their topological charge. In order to detect them we compute the topological charge around a lattice point using the method described in \cite{Achucarro:1999it}.

Finally we track the path of all individual monopoles by taking into account that they cannot travel faster than the speed of light and that they conserve their topological charge. Thus, we can measure the monopole velocity directly using the path they have travelled during the evolution.

\section{Summary}

Here, we have presented a numerical study of semilocal string networks. Firstly, we have discussed the large-scale properties of simulated semilocal networks, covering couplings in the range $0.01 \le \beta \le 0.09$, and damping terms corresponding to expanding universes dominated by radiation and matter. We have demonstrated scaling behaviour for semilocal networks. Then we have described a comparison of our simulations with analytical models: starting from an initial configuration of a semilocal network we are gathering all string segments into length bins and we are comparing the evolution of the segments in each bin using our field theory simulation and our analytic models. An important source of uncertainty in this comparison is related to our lack of knowledge of the velocity in simulations of semilocal strings. Therefore, in the last part we have presented the methods that we will use to compute the transverse string velocity and the monopole velocity. 

\ack
I am supported by the Basque Government grant BFI-2012-228. I acknowledge also financial support from the Basque Government (IT-559-10),the Spanish Ministry (FPA2012-34456), and the Spanish Consolider EPI CSD2010-00064. Additional support for the project was provided by grant PTDC/FIS/111725/2009 (FCT, Portugal).

\end{document}